\documentclass{PoS}

\newcommand{\Dslash}{\ensuremath \raisebox{0.017cm}{\slash}\hspace{-0.2cm} D}
\newcommand{\dslash}{\not{\hbox{\kern-2pt $\partial$}}}

\title{The removal of critical slowing down}

\ShortTitle{The removal of critical slowing down}

\author{\speaker{M. A. Clark}\\
        Center for Computational Science, Boston University, 3
        Cummington St,  MA 02215, USA\\
        E-mail: \email{mikec@bu.edu}}

\author{J. Brannick$^a$, R. C. Brower$^{bc}$,
 S. F. McCormick$^{e}$, T. A. Manteuffel$^{e}$, J. C. Osborn$^{bd}$ and C. Rebbi$^{bc}$\\
  \llap{$^a$} Department of Mathematics, The Pennsylvania State
  University, 221 McAllister Building,
  University Park, PA 16802, USA\\
  \llap{$^b$} Center for Computational Science, Boston University, 3
  Cummington St,
  MA 02215, USA\\
  \llap{$^c$} Department of Physics, Boston University,
  590 Commonwealth Avenue, Boston,  MA 02215,  USA\\
  \llap{$^d$} Argonne Leadership Computing Facility, 9700 S. Cass
  Avenue, Argonne, IL 60439, USA\\
  \llap{$^e$} Department of Applied Mathematics, Campus Box 526, University of
  Colorado at Boulder, Boulder, CO 80309, USA
}

\abstract{We present promising initial results of our adaptive
  multigrid solver developed for application directly to the
  non-Hermitian Wilson-Dirac system in 4 dimensions, as opposed to the
  solver developed in~\cite{Brannick:2007ue}
  for the corresponding normal equations. The key behind the success
  of this algorithm is the use of an adaptive projection onto
  coarse grids that preserves the near null space of the system
  matrix. We demonstrate that the resulting algorithm has weak
  dependence on the gauge coupling and exhibits extremely mild
  critical slowing down in the chiral limit.  }

\FullConference{The XXVI International Symposium on Lattice Field Theory\\
		 July 14-19 2008\\
		 Williamsburg, Virginia, USA}

\begin{document}

\section{Introduction}

We aim to solve the system of equations
\begin{equation}
Dx = b,
\end{equation}
where, in general, \(D = \Dslash + m\) is the Dirac
operator, with \hspace{0.5mm} \(\Dslash\) denoting the tensor product
of the discretized covariant derivatives and the Dirac gamma matrices
and \(m\) the fermion mass.  Herein, we restrict our discussion to the
Wilson-Dirac discretization of \(D\), obtained by using central
covariant differences to discretize $D$ and then adding a properly
scaled second-order Wilson term.  It is well known that, for this
choice, $D$ is a non-Hermitian matrix, with complex eigenvalues that,
in general, satisfy Re(\(\lambda\))\(>0\) and also that $D$ satisfies
a \(\gamma_5\) Hermiticity, i.e., \(D^\dagger = \gamma_5 D \gamma_5\),
and \(H = \gamma_5 D\), where \(H\) is the Hermitian (indefinite)
Dirac operator.  As the fermion mass is decreased
(Re(\(\lambda_{\tiny{\mbox{min}}}\))\(\rightarrow0\) ), $D$ becomes
singular, causing "critical slowing down" (deteriorating convergence)
of the standard Krylov solvers typically used to solve these systems.
Improving convergence of Krylov methods for the Dirac inversion
problem with a suitable preconditioning has been a main topic of
research in lattice QCD for many years and, until more recently, the
cost of inverting $D$ has been prohibitive as this physical light
quark mass is approached.

In the last few years, much progress has been made in addressing this
issue.  Eigenvector deflation \cite{Stathopoulos:2007zi,
  Morgan:2007dq} is a proven technique for accelerating convergence,
provided sufficiently many eigenvectors are used in the deflation;
exact deflation approaches are, however, expected to scale as the
square of the lattice volume \(O(V^2)\) and, thus, become ineffective
for large volume.  An alternative approach is given by the inexact
local mode deflation proposed by L\"{u}scher \cite{Luscher:2007se};
here, only approximate eigenvectors are used in the deflation process
and, due to the local coherence (see below) exhibited by the low
(small magnitude) eigenmodes of the Dirac operator, only a small
number of low mode prototypes are needed for an effective deflation.
Inexact deflation strategies are expected to scale as \(O(V \log V))\)
or even \(O(V)\).

The approach we are proposing is an alternative to deflation, based on
a multigrid (MG) solver for the Dirac operator.  In previous
work~\cite{Brannick:2007ue}, we presented an algorithm for the normal
equations obtained from the Wilson-Dirac system in the context of 2
dimensions, with a \(U(1)\) gauge field.  Here, we extend the approach
directly to the Wilson-Dirac system, which we find is a superior
approach to working with the normal equations, and apply the resulting
algorithm to the full 4-dimensional \(SU(3)\) problem.

\section{Adaptive Multigrid}

The "light" modes, low eigenmodes of the system matrix, typically
cause the poor convergence suffered by standard iterative solvers
(relaxation or Krylov methods); as the operator becomes singular, the
error in the iteratively computed solution quickly becomes dominated
by these modes.  In free field theory, these slow-to-converge modes
are geometrically smooth and, hence, can be well represented on a
coarse grid using fewer degrees of freedom.  Moreover, these smooth
modes on the fine grid now again become rough (high frequency) modes
on the coarse grid.  It is this observation that motivates the
classical geometrical MG approach, in which simple local averaging and
linear interpolation can be used to form corrections to the fine grids
stemming from solutions on a coarse grid.  We hereafter denote the
interpolation operator by \(P\) and restriction operator by \(R\).
Given a Hermitian positive definite (HPD) operator \(A\), taking the
restriction operator as \(R = P^\dagger\) and the coarse-grid operator
as \(A_c = P^\dagger A P\) gives the optimal, in an energy-norm sense,
two-grid correction.  It is natural to recurse upon this approach by
defining the problem on coarser and coarser grids until the degrees of
the freedom have been reduced enough to permit an exact solve.  When
combined with \(m\) and \(n\) relaxation applications before and after
each restriction and prolongation applications, to remove the high
frequency error inherent to that grid, this is known as a
\(V(m,n)\)-cycle, and removes critical slowing down for discretized
PDE problems and scales as \(O(V)\) \cite{Brandt:1977}.

The error propagation operator for the two-grid solver with a single
post-relaxation, with error propagator $S$, is given by
\begin{equation}
E_{TG}=S(I - \pi_A), \qquad\pi_A = P (P^\dagger A P) ^{-1} P^\dagger
A = P A_c^{-1} P^\dagger A.
\end{equation}
Roughly speaking, the performance of a given MG algorithm is related
to \(image\) of \(P\) and how well this approximates the
slow-to-converge modes of the chosen relaxation procedure (called a
``smoother''). More precisely, given $P$ and a convergent smoother,
the two-grid algorithm can be shown to converge provided: for any
fine-grid vector $u \in \mathbb{C}^{n}$, there exists a vector of
coefficients $w_c \in \mathbb{C}^{n_c}$ on the coarse gird such that
the linear combination $Pw_c$, with the columns of $P$ denoting the
coarse-space basis functions, satisfies
\begin{equation}
K(P) = \|A\| \frac{\| u - Pw_c\|}{\|u\|_A} \leq \infty.
\end{equation} 
The above inequality, known as the weak approximation property in the
MG literature, is clearly biased towards the light modes and in
particular suggests that \(P\) approximate eigenvectors with error
proportional to the size of their corresponding eigenvalues.

When solving the Wilson-Dirac system in the interacting theory, the
light modes are not geometrically smooth, and so classical MG, which
assumes the slow-to-converge error is locally constant, fails
completely.  In such settings, where the gauge field is essentially
random and this randomness dictates the local nature of the low modes,
we must alter our definition of the prolongator \(P\) so that it
accurately approximates these light modes.  Because the columns of
\(P\) form the coarse space basis, this is achieved by partitioning
the light modes into subvectors over the aggregates.  The above
approximation property then implies that locally the low modes used in
defining \(P\) must form a basis for the low modes of the system matrix,
which for most simple pointwise smoothers are also the low modes not
effectively treated by the relaxation method.  In the physics
community, this notion that a small set of vectors partitioned into
local basis functions can approximate the entire lower end of the
eigenspectrum of a matrix is known as local coherence.  It is this
idea of local coherence that leads to the success of
L\"{u}scher's~\cite{Luscher:2007se} deflation approach as well as our
MG solver.

With a convergent MG algorithm defined, we are now left with the
problem of adaptively finding a representation of the light modes to
allow us to define \(P\).  One viable approach, known in the MG
literature as adaptive smooth aggregation (\(\alpha\)SA)
~\cite{Brezina:2004} , is given by iteratively computing the low modes
and then adjusting \(P\) to fit these computed modes.  The general
algorithm for computing these prototypes for some matrix \(A\)
proceeds as follows:
\begin{enumerate}
\item Apply the MG relaxation to the system \(Ax=0\), where \(x_0\)
  is taken to be a random vector.
\item Terminate the solver after \(l\) iterations.  The
  current iterate \(x_l = -e_l = v_1\) will be a representation of the slow
  to converge components.
\item Decompose this error vector into aggregates (blocks).  The
  components within each of these blocks represent a column of the
  prolongator \(P\).
\end{enumerate}

Numerical experience suggests that for the 4-dimensional Wilson-Dirac
system with disordered background gauge field, more than a single near
null space vector is needed in defining a \(P\) operator that
sufficiently captures the near null space of the operator in question,
that is, more than a single vector is needed to ensure local
coherence.  In the above discussion, the initial prototype of the null
space used in defining \(P\) is computed using relaxation.  More
generally, the current MG solver is applied to \(Ax =0\) to generate
prototypes of the algebraically smooth error, yielding at each
adaptive step the matrix \(V^k = [ v_1, ..., v_k]\), with the
\(v_i\)'s denoting the computed candidate vectors approximating the
near kernel of the operator \(A\), where \(V^k\) is augmented until
convergence is deemed sufficient, say for $k=N_v$ candidate vectors.
At each stage in which \(V^k\) is augmented, one defines the
(tentative) prolongation operator \(P\) and the representation of
\(V^k\) on the coarse lattice \(V^k_c\) in the usual way, that is, by
enforcing the following relations:
\begin{equation}
P V^k_c = V^k \hspace{10mm}\mbox{and}\hspace{10mm} P^{\dagger} P = I,
\end{equation}
that is to say one performs a \(QR\) decomposition on the partitioned
candidate vectors, where \(Q\) represents the columns of prolongator
and \(R\) represents the coefficients in the coarse basis, i.e.,
\(V^k_c\).

\section{Formulating an algorithm}

The original adaptive smoothed aggregation approach introduced in
\cite{Brezina:2004} is essentially a black-box method, where the
blocking strategy is chosen using an algebraic strength-of-connection
measure.  In lattice QCD, the system is discretized on an uniform
hypercubic lattice, with unitary connections link the sites.  With
this in mind, our blocking strategy is to use a regular geometric
structure, e.g., \(4^d\) geometric blocking.

In solving the Wilson-Dirac system, we consider two
approaches:
\begin{enumerate}
\item Apply the adaptive MG approach to the normal equations.
\item Formulate a MG algorithm directly for the Wilson-Dirac
  operator.
\end{enumerate}

The normal equation approach has the obvious advantage that the
operator in question is HPD; hence, the standard Galerkin operator
definition \((D^\dagger D)_c = P^\dagger (D^\dagger D) P\) is optimal
and all MG convergence proofs can be applied directly.  However, the
condition number of \(D^\dagger D = H^2\) is the square of that of
\(H\) and the coarse operator cannot be written as the product of
nearest neighbour couplings, leading to loss in operator sparsity.

For non-HPD systems, the usual MG convergence proofs generally do not
apply.  However, in this case, a significant amount of insight is
obtained by considering the spectral decomposition of \(D =
|\psi_\lambda \rangle \lambda \langle \tilde\psi_\lambda| \), where
\(\psi\) and \(\tilde\psi\) are the right and left eigenvectors
respectively, both having eigenvalue \(\lambda\).  If we now consider
eigenvector deflation, in which case we use a Petrov-Galerkin oblique
projection, i.e., treat the left and right spaces separately, to
remove a given eigenvalue \(\lambda\), then we have:
\begin{equation}
\mathcal{P} = \left(1 - D|\psi_\lambda\rangle\frac{1}{\lambda} \langle
  \hat{\psi}_\lambda| \right) =
\left(1-D|\psi_\lambda\rangle\langle\hat{\psi}_{\lambda'}|D|\psi_\lambda\rangle^{-1}
  \langle\hat{\psi}_{\lambda'}|\right) = \left(1 - D P (R D P)^{-1} R
\right).
\end{equation}
We thus see that prolongation should be defined using ``right null
space vectors'' and restriction from ``left null space vectors''.
Naively, this suggests that we define prolongation using smoothed
vectors of \(D\) and restriction from smoothed vectors of
\(D^\dagger\).  However, because of the \(\gamma_5\) symmetry of the
Wilson-Dirac operator, we have \(\psi_\lambda =
\gamma_5\tilde{\psi}_{\lambda^*}\); and hence, a vector rich in low
right eigenvectors can be converted to one rich in low left
eigenvectors by simple multiplication by \(\gamma_5\).  Given the
current residual \(r_0\), our coarse-grid correction is thus given by
\begin{equation}
x_c = \alpha\, P (P^\dagger \gamma_5 D P)^{-1} P^\dagger \gamma_5 r_0 = \alpha\,P
(P^\dagger H P)^{-1} P^\dagger \gamma_5 r_0,
\end{equation}
where the step-size \(\alpha\) is defined below.  If we block the spin
dimension, however, it is possible for the coarse operator to have
exactly zero eigenvalues\footnote{Take for example the free field
  operator, where the null space vector is given by the constant.
  Thus, \(P^\dagger \gamma_5 P = 0\), and our coarse-grid correction
  is ill-defined.}.  This is easy to avoid by keeping chirality
intact, i.e., \([\gamma_5,P] = 0\).  In this way, the \(\gamma_5\)
factors cancel out in the overall coarse-grid correction, yielding
the former "naive" result \(R = P^\dagger\).

Convergence of our coarse-grid correction is not guaranteed: if we
perform an eigenvector expansion of the low modes used to define the
prolongator 
\begin{equation}
  v_i = \sum_{\lambda, \lambda^*} \langle \psi_\lambda |
  v_i \rangle \psi_{\lambda} + \langle \psi_{\lambda^*} | v_i \rangle
  \psi_{\lambda^*} = \sum_{\lambda, \lambda^*} c_\lambda \psi_{\lambda}
  + c_{\lambda^*} \psi_{\lambda^*},\end{equation}
the left space is given by 
\begin{equation}
  \hat{v}_i = \sum_{\lambda, \lambda^*} c_\lambda
  \hat{\psi}_{\lambda^*} + c_{\lambda^*} \hat{\psi}_{\lambda}.
\end{equation}
Since in general \(c_\lambda \neq c_{\lambda^*}\), the left and right
spaces are not equivalent, and our coarse-grid correction can be
divergent.  Convergence can be imposed by choosing the step-size
\(\alpha\) to always minimize the resulting residual \(r_{1} = r_0 - D
x_c\), i.e.,
\begin{equation}
  \alpha = \frac{\langle D x_c | r_0 \rangle }{ \langle D x_c | D x_c
    \rangle }.
\end{equation}
With the prolongator and coarse-grid operator defined,
all that remains is to define a suitable relaxation procedure that
effectively damps the eigenvectors of the system matrix with
eigenvalues that are large in magnitude.  Classical MG methods use
either Jacobi or Gauss-Seidel smoothing, which are either inefficent
or cannot be applied directly to non-HPD operators.  Currently, we use
an under-relaxed minimal residual smoother (with under-relaxation
parameter \(\omega = 0.8\)).  This yields a simple parallel approach
that reduces the residual in the \(D^\dagger D\) norm, ensuring that
error components corresponding to eigenvectors with large eigenvalues
are damped quickly.

\section{2d Results}

\begin{center}
  \begin{figure}
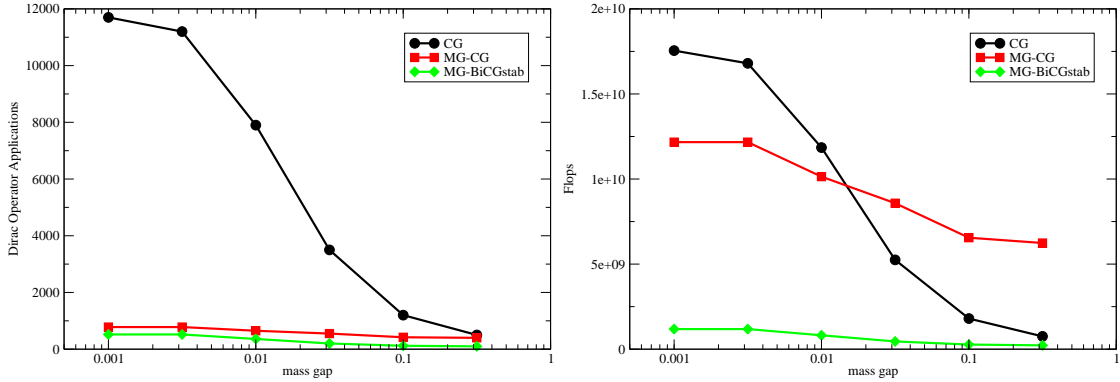

    \includegraphics[height=5cm]{2d_dirac}
    \includegraphics[height=5cm]{2d_flops}
    \caption{Comparison of CG, MG-CG (\(N_v = 8\)) and MG-BiCGstab
      (\(N_v = 3\)) algorithms: left panel compares total number of
      Wilson matrix-vector operations; right panel compares actual
      flops (\(V=128^2\), \(\beta = 6.0\)). }
    \label{fig:2d}
  \end{figure}
\end{center}

Before presenting results of our solver to the full \(4d\) problem, we
compare the normal equations and direct approaches in the \(2d\) QED
context.  Here, a 3-level \(V(2,2)\)-cycle approach has been used,
with \(4^2\) geometric blocking and a machine precision solve has been
used on the coarsest (\(V=8^2\)) lattice.  The normal equation
approach is used to precondition CG (MG-CG), and the direct approach
to precondition BiCGstab (MG-BiCGstab).  In the left plot of figure
\ref{fig:2d}, which compares the total number of Dirac operator
applications, we see that both approaches effectively remove critical
slowing down, with both methods leading to around an order of
magnitude reduction of operator applications.  Looking at the actual
number of floating point operations (right panel), it is seen that
MG-BiCGstab is around an order of magnitude cheaper than MG-CG as the
chiral limit is taken, indeed, MG-CG is not competitive with regular
CG except at very small mass.  This large disparity in cost between
the two approaches is due to the increased operator complexity for the
coarse normal equation operator (\(N_v = 8\)) compared to the direct
approach (\(N_v = 3\)).  It should be noted that we have tested the
validity of our algorithm in the range \(\beta=0.1\ldots100\) and
found very little dependence with the resulting MG performance.

\vspace{-13mm}
\section{4d Results}
\begin{center}
  \begin{figure}
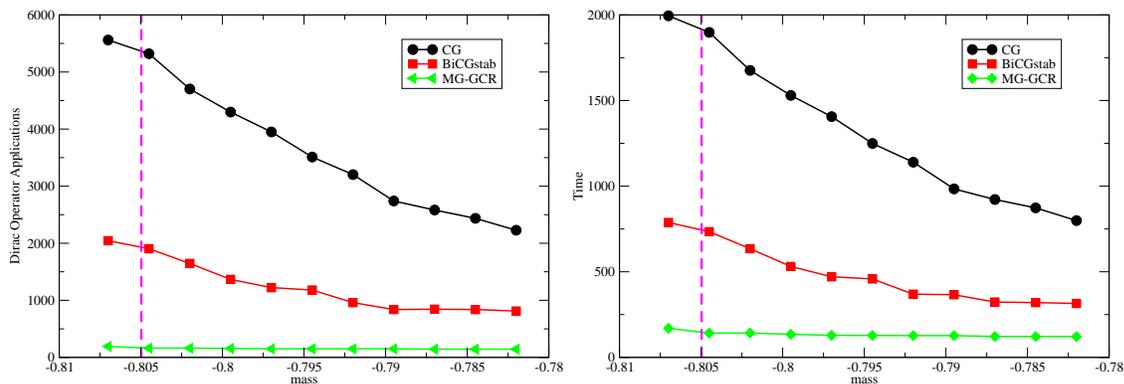

    \includegraphics[height=5cm]{16_32}
    \includegraphics[height=5cm]{time}
    \caption{Comparison of CG, BiCGstab and MG-GCR algorithms, left
      panel compares total number of Wilson matrix-vector operations,
      right panel compares time to solution (\(V=16^3.32\), \(\beta =
      6.0\), \(m_{crit} = -0.8049\), \(N_v = 20\)). }
    \label{fig:4d}
  \end{figure}
\end{center}

Given the \(2d\) results, we applied only the direct approach to the
\(4d\) operator; here we found that \(N_v = 20\) vectors were required
to sufficiently capture the null space of the Dirac operator (this
matches L\"{u}sher's result in \cite{Luscher:2007se}).  The outer
Krylov solver was switched to GCR(50) since our MG operator is a
non-stationary solver (this leads to breakdown in the recursion
relations of non-restarted solvers).  In figure \ref{fig:4d}, we
demonstrate the benefit of our MG-GCR algorithm over CG and BiCGstab,
where our MG algorithm is a modified 3-level \(V(2,2)\)-cycle: between
each restriction and prolongation application on the fine grid, there
are 4 applications of the next coarsest level V-cycle, which ensures
that much more accurate solutions are obtained to the first coarse-grid.
Again, \(4^d\) geometric blocking is used and, in moving from the
original fine grid to the first coarse grid, we also include colour and
spin components with like chirality in the block (but not the full
spin dimension).  On the left panel, we plot the total number of Dirac
operator applications until convergence, where we see that there is an
order of magnitude improvement over BiCGstab.  On the right panel,
comparing time to convergence, we see that the true benefit is closer
to a factor of 4-5 over BiCGstab as the chiral limit is taken.

\vspace{-10mm}
\section{Conclusion}
In this work, we introduced a new multigrid algorithm and showed that
it removes critical slowing down as the quark mass is taken to zero in
the Wilson-Dirac operator.  Future work in this area will focus on
further refining our algorithm to reduce the setup cost of the
algorithm, i.e., attempting to reduce \(N_v\) from 20 and applying
these techniques to staggered and chiral fermions.

\vspace{5mm} 

{ \small This research was supported under: DOE grants
  DE-FG02-91ER40676, DE-FC02-06ER41440, DE-FG02-03ER25574 and
  DE-FC02-06ER25784; Lawrence Livermore National Laboratory contracts
  B568677, B574163 and B568399; and NSF grants PHY-0427646, DMS-0749317,
  0810982 and 0749202. }

\end{document}